\documentclass{aastex}
\usepackage{emulateapj5}

\shorttitle{Spectral Variations in Early-Type Galaxies}
\shortauthors{Concannon, Rose \& Caldwell}

\begin{document}

\title{Spectral Variations in Early-Type Galaxies as a Function of Mass}

\author{Kristi Dendy Concannon and James A. Rose}
\affil{Department of Physics and Astronomy, CB \#3255, University of North Carolina,
       Chapel Hill, NC 27599}
\email{dendy@physics.unc.edu, jim@physics.unc.edu}

\and

\author{Nelson Caldwell}
\affil{F. L. Whipple Observatory, Smithsonian Institution, 
       P.O. Box 97, Amado, AZ 85645}
\email{caldwell@flwo99.sao.arizona.edu}

\begin{abstract}

We report on the strengths of three spectral indicators
--- Mg$_2$, H$\beta$, and Hn/Fe --- in the integrated light of
a sample of 100 field and cluster E/S0 galaxies.  The measured indices are sensitive
to age and/or and metallicity variations within the galaxy sample. Using linear
regression analysis for data with non-uniform errors,
we determine the intrinsic scatter present among the spectral indices of our
galaxy sample as a function of internal velocity dispersion.  
Our analysis indicates that there is significantly more intrinsic scatter
in the two Balmer line indices than in the Mg$_2$ index, indicating that the Balmer 
indices provide more dynamic range in determining the age of a 
stellar population than does the Mg$_2$ index.  Furthermore, the scatter is much larger for the low
velocity dispersion galaxies, indicating that star formation has occurred more
recently in the lower mass galaxies.

\end{abstract}


\section{Introduction}

An enormous effort has been made in recent years 
to disentangle the degenerate effects of age and metallicity on the integrated
light of a galaxy.  This age-metallicity degeneracy (Worthey 1994, 1996)
plagues both the photometric colors and the integrated spectrum of a galaxy,
making it extremely difficult to separate the effects of the two. 
Because broadband colors cannot break the degeneracy (Bica et al.\ 1990; Charlot \& Silk 1994),
the best way to determine a unique solution to the problem is to use a 
combination of population synthesis models and spectral line indices. 
Absolute ages and metallicities determined using this method are 
naturally subject to systematic errors, yet
relative values are still of great interest and are generally more reliable (e.g., Trager et al.\ 2000a).
Although no single spectral index can completely separate the effects of
age and metallicity, it is possible to determine the relative range empirically
by plotting a single spectral index
with a predominant age or metallicity sensitivity versus a mass 
indicative parameter such as the absolute magnitude or internal
stellar velocity dispersion.  The spread in index values which is 
observed as a function of mass is then directly related to the range
in age and/or metallicity in the galaxy sample.  

The most widely used spectral index system is the set of indices originally
defined in Burstein et al.\ (1984) and further refined in Worthey et al.\ (1994)
and Trager et al.\ (1998), now 
commonly known as the Lick system.  In this system, the
H$\beta$ index is used as the primary age-sensitive spectral indicator,
whereas the Mg and Fe indices are used as the primary metallicity indicators.
Using the Lick indices, many groups have been able to place constraints
on the ages and metallicities of various galaxy samples.
For example, Trager et al.\ (2000a) find that for a sample of early-type galaxies in 
low density environments there is a large spread in H$\beta$ values (i.e., age), but
very little variation in metallicity.  For galaxies in the Fornax cluster, though,
Kuntschner (2000) finds the opposite effect, in that a large spread in metallicity
is present with little variation in age.  Whether such contrasting results
are a product of the different environments in the two samples is an interesting
and still unanswered question.  It is clear, however, that a comparison of the range in age and 
metallicity of galaxies in different environments will lead to a better understanding of the 
parameters which govern the evolution of galaxies.

To constrain galaxy evolution scenarios,
it is imperative that we first derive reliable values for the ages and 
metallicities of galaxies. In this paper we present first results of a sample of 100 early-type 
galaxies covering a large baseline in mass and a large range in wavelength in order 
to use both the Lick/IDS age/metallicity
indicators as well as additional indices centered around the higher order Balmer lines.
Although the data will eventually be used to determine
absolute ages and metallicities, and thereby place constraints on the theories
governing galaxy formation and evolution, these goals are beyond the scope of this paper.
In this letter, we demonstrate the importance of 
the lower mass galaxies and higher order Balmer lines for obtaining a complete
view of the ages of early-type galaxies.  The paper is organized such that
the data and reduction procedures are summarized in \S 2, the spectral 
indices and the index diagrams are presented in \S 3, and the analysis of these
diagrams is given in \S 4 and discussed in \S 5.

\section{Observations}

Our galaxy sample consists of 100 early-type galaxies, 70 of which are 
field galaxies and 30 of which are in the Virgo cluster.  The field 
galaxies are drawn from the CfA redshift survey (Huchra et al.\ 1983); the cluster galaxies 
are from the Virgo Cluster Catalog (VCC) of Binggeli et al.\ (1985).  
The observed galaxies cover a range in absolute magnitude of $-16 < M_B < -22$
and have been previously classified as either E or S0.

Long slit spectra of the galaxies were obtained at the F.\ L.\ Whipple 
Observatory with the 60" Tillinghast telescope, the FAST spectrograph and 
a Loral 512 x 2688 pixel CCD (Fabricant et al.\ 1998) during seven different
observing runs between 1997 September and 1999 October
in seeing conditions of 1$\arcsec$ or better.  
A 3$\arcsec$ slit and the 600 line/mm grating were used to give a dispersion
of 0.75\AA/pixel and 3\AA\ FWHM spectral resolution and a spectral 
coverage of 3500\AA\ - 5500\AA.  Each exposure was 
1800 seconds with 2-4 exposures taken for each galaxy. The slit was oriented at
the parallactic angle so as to preserve the blue light, resulting in a S/N per pixel
between 70 and 100 at 4000\AA\ in the final coadded spectrum. 

The spectra were reduced in IRAF with standard reduction procedures including
bias and dark subtraction as well as flat fielding.  The spectra
were wavelength calibrated using an internal HeNeAr lamp and were
flux calibrated using standard stars from the Massey et al.\ (1988) spectrophotometric
catalog. Multiple galaxy exposures were added together and all
galaxies were de-redshifted to the same rest wavelength.  Velocity dispersions were
determined using the Fourier cross-correlation method (e.g., Tonry \& Davis 1979) 
in the task \emph{fxcor} and various stellar templates which were observed during twilight. 
All spectra were then smoothed to the same effective dispersion of 230 km/s 
to insure that each object experienced the same intrinsic broadening. 

\section{Spectral Indices}

Two types of spectral indices were measured from the calibrated and smoothed spectra,
the pseudo-equivalent width indices of the Lick group (Burstein et al.\ 1984; Worthey et al.\ 1994)
and the line ratios of Rose (1994; Caldwell \& Rose 1998).
The Lick/IDS family of indices measures the pseudo-equivalent width of
absorption line features such as H$\beta$ and the Mg bands which lie redward of 4000\AA.
The Lick equivalent width indices are measured by centering a bandpass of typically 30-40\AA\ on the
absorption feature and using a red and blue sideband of similar width as the continuum 
reference.  In addition to the Lick system, we have also chosen to work at 
bluer wavelengths where young stars contribute a larger fraction of the light.
The indices used in this region are the Rose spectral line ratios,
which are defined as the ratio of the residual intensity in two neighboring spectral lines.
For our galaxy sample, we have measured a total of 36 spectral indices, 
22 Lick indices and 14 Rose indices,
and report in this letter on three of the key age/metallicity diagnostics. 
The spectral indices analyzed here are: Mg$_2$, a Lick index which measures 
the strength of the Mg~H + Mg~b 
molecular bands, H$\beta$, another Lick index which
measures the equivalent width of the
H$\beta$ feature at 4861\AA, 
and Hn/Fe, a Rose spectral line index which is an average of three hydrogen
to iron line ratios\footnote{$Hn/Fe = <H\delta/Fe4045 + H\gamma/Fe4325 + H8/Fe3859>$} 
and is defined such that a lower index value corresponds to a stronger Balmer line 
strength.  

For each index, the error for an individual galaxy is determined from 
the rms scatter in the repeat measurements.  Since this method is unreliable 
for those galaxies which have only two observations,
we have defined our errors for a particular index and galaxy
utilizing the errors in all indices and galaxies in the following manner.  The
error, $\Lambda_{ij}$, in index \emph{j} for galaxy \emph{i} 
(where there are a total of \emph{m} indices and \emph{n} galaxies)
is given by:
\begin{equation}
\Lambda_{ij} = \frac{\sum_{i=1}^n \varepsilon_{ij} \sum_{j=1}^m \varepsilon_{ij}}{\sum_{i,j=1}^{m,n} \varepsilon_{ij}}
\end{equation}
where $\sum_{i=1}^n \varepsilon_{ij}$ is the rms error in the $j^{th}$ index averaged over all galaxies,
$\sum_{j=1}^m \varepsilon_{ij}$ is the error in the $i^{th}$ 
galaxy averaged over all indices of the same type (i.e, all Lick indices or all Rose indices),
and $\sum_{i,j=1}^{n,m} \varepsilon_{ij}$ is the average error in all indices of the same type for all 
galaxies. The average fractional uncertainty in each index is 1\% for Hn/Fe and Mg$_2$, 3\% for H$\beta$
and 2\% for the velocity dispersion measurements.  

The three indices reported in this paper are shown in Figure~1 as
a function of log $\sigma$ (i.e., as a function of mass). 
The line is the best fit to the high-mass data extended to the low-mass end;
an explanation of this fit is given in \S 4.
Comparing these diagrams, two effects are immediately evident.  First, the scatter 
among galaxies appears to increase between the high-mass end and the low-mass end of the diagrams.
Second, this increase in scatter is significantly more dramatic in both the H$\beta$ and Hn/Fe
diagrams than it is in the Mg$_2$ diagram. 
Some of the observed scatter of the galaxies about the linear fits is obviously due to measurement
errors associated with the data, but variations in the age and metallicity of the
galaxies play a contributing role as well, and in practice, it is difficult to separate this
intrinsic scatter from the observational errors.  In the following section, we describe a
statistical analysis of the scatter in the index diagrams and discuss their implications.

\section{Results}

In each diagram in Figure~1, a linear relation is readily identifiable for
the galaxies with log $\sigma > 2.0$.  This property is related to the well-known
color-magnitude relation of elliptical galaxies (Visvanathan \& Sandage 1977)
and is thought to be driven by metallicity (Faber 1973, 1977). 
The relation is naturally explained by the supernova-driven wind model (Larson 1974; 
Arimoto \& Yoshii 1987) 
in which the more massive galaxies are better able to retain
their supernova ejecta than the smaller galaxies and thus become more
metal-rich and redder.  Because the high-mass
galaxies represent a fiducial, normal sample of galaxies, we 
used only these points to determine the best fit to the data and extended the fit
to the lower mass galaxies. The three outlying points above log $\sigma=2.0$
were not used in the regression analysis, since each of these galaxies is known to be in 
an interacting system with star formation, possibly triggered by this interaction.
The least-squares fitting is done using the BCES (bivariate correlated errors and intrinsic scatter)
method of Akritas \& Bershady (1996)
that fits data for which there is some intrinsic scatter present and
for which the error bars are heteroscedastic (non-uniform) in both the \emph{x} and \emph{y}
coordinates. Because we are interested in measuring the intrinsic scatter as a function of mass
(i.e., internal velocity dispersion), the chosen fit minimizes the residuals in the 
indices (\emph{y}-variables), which have, on
average, larger errors than the velocity dispersion (\emph{x}-variable).
To verify the fits, the analysis was repeated for 1000 bootstrap re-samplings 
of the data. We find that the BCES and bootstrap fits agree within their associated errors. 

To separate the intrinsic scatter, $\eta$, from the scatter due to measurement errors,
we used an estimate given by Fioc \& Rocca-Volmerange (1999):
\begin{equation}
\eta^2=\sum[(y_i - Bx_i - A)^2 - (B^2\xi_{xi}^2 + \xi_{yi}^2)]
\end{equation}
where \emph{B} and \emph{A} are the slope and intercept respectively of the BCES
regression analysis, $x_i$ is the observed log $\sigma$, $y_i$ is 
the observed index value
and $\xi_{xi}$, $\xi_{yi}$
are the associated errors in the $x_i$ and $y_i$ observations.  
The galaxies were divided into two bins, log $\sigma > 2.0$ and log $\sigma < 2.0$,
the scatter was calculated for each subsample, and the analysis was
verified with Monte-Carlo simulations. 
The measured intrinsic scatter for both the high- and low-mass subsamples is shown
as the vertical dotted lines in the lower right corners of Figure~1.  

The results of the statistical analysis of the three spectral indices, H$\beta$, Mg$_2$, and Hn/Fe,
are shown in Table~1.  Here, the intrinsic scatter in each subsample is denoted by 
$\eta$; the error in $\eta$, estimated from bootstrap re-samplings, is 
given by $\delta_\eta$. 
The reader should be aware that because the two Lick
indices and the Rose index are defined differently (as equivalent widths and a line
ratio, respectively) it is not meaningful to compare the absolute difference in the
intrinsic scatter between two indices, i.e., it is only meaningful to compare relative
differences in the same index.  The quantitative analysis clearly verifies our previous predictions
in that a non-zero intrinsic scatter exists in each of the index diagrams and
that this scatter is greater for the low-mass galaxies than for the high-mass galaxies.
In fact, the intrinsic scatter increases by a factor of 1.5 from the low-mass to the
high mass galaxies in the Mg$_2$ diagram, by a factor of 2.6 in the H$\beta$ diagram,
and by a factor of 5.2 in the Hn/Fe diagram (R value in Table~1).  In addition to the
three indices reported here, we have also estimated the intrinsic scatter in another
Lick index, Mg~b, which has become a popular age/metallicity indicator (e.g, Trager et al.\ 2000a).
For this index, we find that the intrinsic scatter increases by a factor of 2.2 from the
high-mass to the low-mass galaxies.  It is worth noting that this value is between the estimates
for Mg$_2$ and H$\beta$.

From this analysis, it is evident that more
intrinsic scatter exists in the Balmer indices than in the Mg$_2$ index. 
If the factor driving the scatter were metallicity instead of age, we would actually
expect to see a larger scatter in the Mg$_2$ data than in the Balmer indices,
since Mg$_2$ is more sensitive to metallicity than to age in comparison with H$\beta$ and other
Balmer lines (e.g., Buzzoni et al.\ 1994).  However, the
opposite situation is observed.  Because
we see more variations in the Balmer indices than the Mg$_2$ index, especially at low mass,
the index diagrams imply that the lower-mass galaxies exhibit a larger range in ages than
the high-mass galaxies.

\section{Discussion}

The results of this analysis bring forth two key issues that must be addressed.
First, since the two Balmer line indices exhibit more intrinsic scatter than the
Mg$_2$ index, this qualifies them as more robust age indicators.  The relatively
little intrinsic scatter observed in the Mg$_2$-log $\sigma$ diagram indicates that 
this index is not well suited for constraining the age of a stellar population.
The fact that significant age variations may be hidden in the Mg$_2$-log $\sigma$ relation 
has been previously noted by both Worthey et al. (1996) and J$\o$rgensen (1999) and
has been studied in detail by Trager et al.\ (2000b).
This finding may have a significant impact on the previous results of authors
who have used the Mg$_2$ index as their principal age diagnostic (e.g., Bernardi et al.\ 1998),
and their conclusions may need to
be re-evaluated using a more age-sensitive index such as H$\beta$ or Hn/Fe.

The second key result is the amount of scatter observed for the low-mass galaxies
in the index diagrams.  The dramatic increase in scatter, particularly at the
low-mass end of the Balmer line diagrams, indicates that the smaller galaxies
have experienced a more varied star formation history and that these galaxies have
a larger range in age than their brighter counterparts.  This finding places
significant constraints on a hierarchical formation scenario such as the CDM model
(Baugh, Cole \& Frenk 1996; Kauffmann 1996), since in this case 
one would naively expect to see that low-mass galaxies are
older than the larger galaxies (Kauffmann \& Charlot 1998). However, because
the majority of the low-mass galaxies in our sample are in the Virgo cluster,
it is difficult to determine whether the observed variation in age is simply a product of 
the different environments.  Even if this were the case, though, in a hierarchical scenario
we would expect that galaxies in clusters would be older than field galaxies, since merging
will happen more rapidly in clusters.  Thus, regardless of the environmental influence, our
results appear to place significant constraints on the hierarchical picture.

In a forthcoming paper, we will report on use of the higher order Balmer index and
population synthesis models to determine the spread of age
and metallicity of our sample, and thereby place quantitative constraints on formation and
evolution scenarios.




\acknowledgments

We would like to thank to William J.\ Thompson for much valuable advice and 
discussion concerning the statistical analysis in this paper.  Thanks are also due
to Michael Akritas and the 
SCCA\footnote{http://www.stat.psu.edu/$\sim$mga/scca/} for their advice.
This research was supported
in part by a North Carolina NASA Space Grant Consortium Graduate Fellowship
and by NSF grant AST-9900720 to 
the University of North Carolina.





\clearpage



\begin{figure}[p]
\plotone{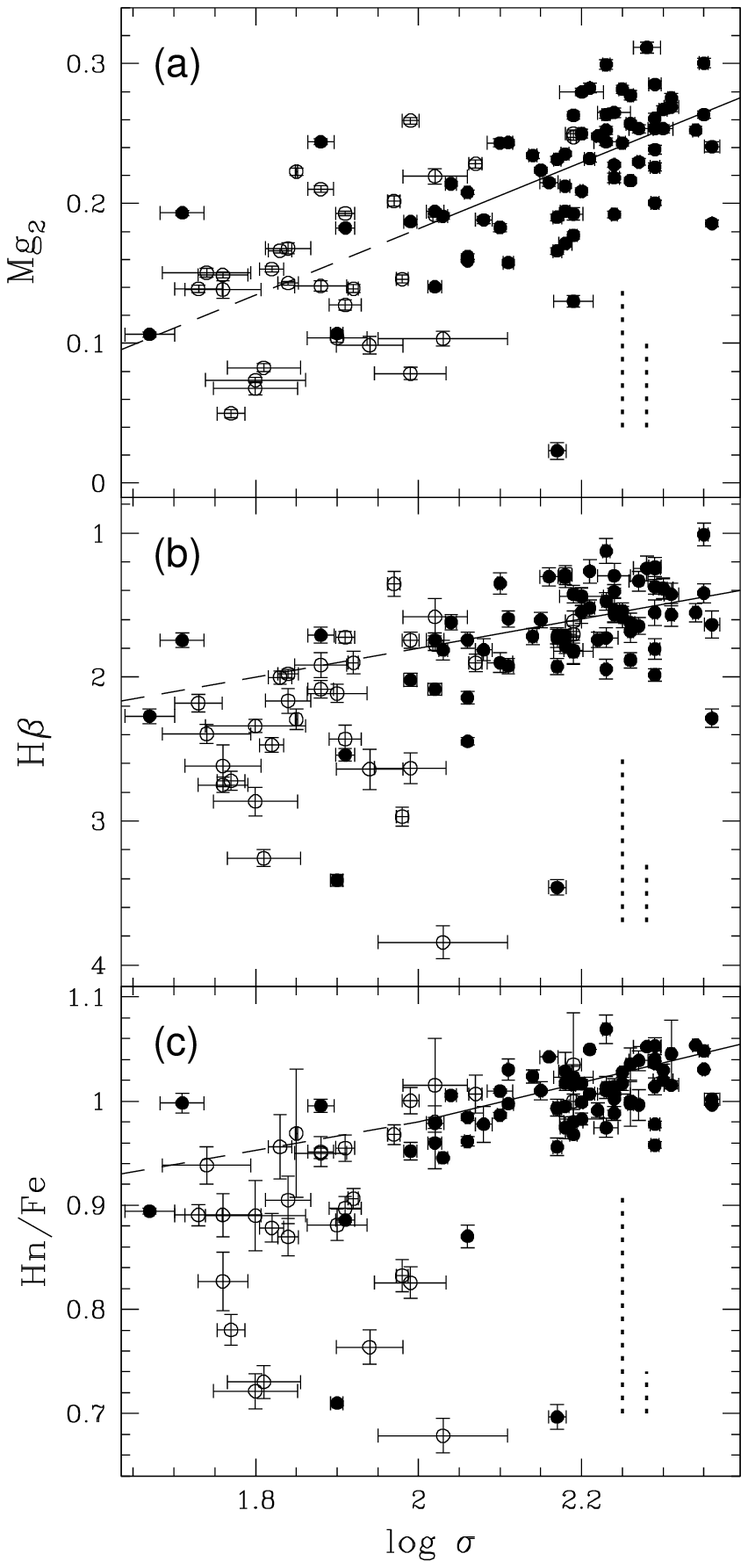}
\figcaption{Three spectral indices (a) Mg$_2$, (b) H$\beta$, (c) Hn/Fe
plotted versus log $\sigma$.  The solid
circles are field galaxies; the open circles are Virgo cluster galaxies.  The
regression line is the calculated fit with the dashed portion indicating 
the region of the line which is extrapolated. The vertical dotted lines 
represent the measured amount of intrinsic scatter in each subsample; the
high-mass scatter is on the right, the low-mass scatter is on the left. 
The index values in (a) are in magnitudes; the values in
(b) are equivalent widths in Angstroms; the index in (c) is a ratio.}
\end{figure}

\clearpage
\begin{deluxetable}{crrrrrr}
\footnotesize
\tablecaption{Measured Intrinsic Scatter in Index Diagrams \label{tbl-1}}
\tablewidth{0pt}
\tablehead{
\colhead{Index} & \colhead{Subsample\tablenotemark{a}}   & \colhead{$\eta$}   &
\colhead{$\delta_\eta$} &
\colhead{R\tablenotemark{b}} & $\delta_{R}\tablenotemark{c}$
}
\startdata
Mg$_2$ & low mass & 0.050 & 0.005 & 1.52 & 0.21 &\\
       & high mass & 0.033 & 0.003 &   &\\
H$\beta$ & low mass & 0.578 & 0.084 & 2.60 & 0.48 &\\
       & high mass & 0.222 & 0.025 &  &\\
Hn/Fe  & low mass & 0.105 & 0.015 & 5.25 & 1.01 &\\
       & high mass & 0.020 & 0.003 &    &\\
\enddata

\tablenotetext{a}{Low mass: log $\sigma <$ 2.0; high mass: log $\sigma >$ 2.0}
\tablenotetext{b}{Ratio of intrinsic scatter in low mass galaxies to that
in high mass galaxies}
\tablenotetext{c}{Error in the intrinsic scatter ratio}
\end{deluxetable}




\end{document}